\documentstyle[tighten,preprint,epsfig,amsfonts,prd,aps]{revtex}
\begin{document}

\draft
\preprint{DAMTP-2000-11, grqc/0001070}

\title{Coordinate singularities in harmonically-sliced cosmologies}
\author{Simon D. Hern}
\address{Department of Applied Mathematics and Theoretical Physics,\\
         Silver Street, Cambridge, CB3 9EW, England \\
         Email: S.D.Hern@damtp.cam.ac.uk}
\date{23rd January 2000}

\maketitle

\begin{abstract}
Harmonic slicing has in recent years become a standard way of
prescribing the lapse function in numerical simulations of general
relativity. 
However, as was first noticed by 
Alcubierre~[Phys.\ Rev.~D {\bf 55}, 5981 (1997)], 
numerical solutions generated using this slicing condition can show
pathological behaviour.
In this paper, analytic and numerical methods are used to examine
harmonic slicings of Kasner and Gowdy cosmological spacetimes.
It is shown that in general the slicings are prevented from covering
the whole of the spacetimes by the appearance of coordinate
singularities.
As well as limiting the maximum running times of numerical simulations,
the coordinate singularities can lead to features being produced
in numerically evolved solutions which must be distinguished
from genuine physical effects.
\end{abstract}
\pacs{04.25.Dm, 98.80.Hw}

\input epsf.tex
\def\DrawFig#1{} 

\narrowtext

\section{Introduction}
\label{sec.Intro}
In the 3+1~formulation of general relativity, the freedom
in choosing a coordinate system through which to describe a spacetime
is replaced by freedom in choosing two gauge quantities: 
the {\it lapse function}~$N$, which controls the slicing of the
spacetime into a foliation of
spatial hypersurfaces, and the {\it shift vector}~$N^i$,
which defines a set of reference world lines threading those spatial
slices. 
Numerical simulations based on the 3+1~formulation are
required to  prescribe a method for evaluating the lapse and the shift
in terms of the geometrical quantities 
(the {\it intrinsic metric}~$h_{ij}$ and 
the {\it extrinsic curvature}~$K_{ij}$) and the matter
fields (the density~$\rho$, momentum~$J^i$, and 
stress~$S^{ij}$) that are known on each spatial slice.
The use of a {\it harmonic slicing condition\/} to determine the
lapse function is appealing for both practical and theoretical
reasons, and the incorporation of such a slicing condition in
numerical codes is becoming increasingly common.
However, recent work (Alcubierre and Mass\'o~\cite{Alcu97,AlMa98},
Geyer and Herold~\cite{GeHe97}) has found evidence that the use of
harmonic slicing can produce pathological behaviour in numerically
evolved spacetimes.

This paper addresses the question as to how suitable harmonic
slicing is for work in numerical relativity.
In section~\ref{sec.Review} the main features of the harmonic slicing
condition are reviewed.
Following an approach similar to that of Geyer and 
Herold~\cite{GeHe97,GeHe95} the behaviour of `planar' harmonic
slicings of the Minkowski spacetime is analysed in section~\ref{sec.Mink};
it is demonstrated there that such slicings always cover the whole of
the spacetime.
A similar statement does not however hold true for the Kasner
spacetime.
Results derived in section~\ref{sec.Kasner} show
that coordinate singularities of the type found by
Alcubierre~\cite{Alcu97} 
appear in harmonic slicings of the Kasner spacetime under reasonably
general conditions.
Section~\ref{sec.Gowdy} shows that, furthermore, the results for the
Kasner spacetime carry over directly to harmonic slicings of the
more general class of Gowdy~$T^3$ spacetimes.
In section~\ref{sec.Numeric} results are presented from numerical
simulations of the Kasner spacetime which use harmonic slicing.
The coordinate singularities predicted by the analysis of earlier
sections are indeed encountered, and the
behaviour of the numerical solutions at times just prior to the
formation of coordinate singularities is examined.
Section~\ref{sec.Discuss} concludes the discussion by considering the
implications that these results have for the use of harmonic slicing in
numerical relativity.

\section{Harmonic Slicing in Numerical Relativity}
\label{sec.Review}
When considering Einstein's equations as a 3+1~evolution
system, a spacetime is described in terms of a foliation
$\{ \Sigma_t : t \in {\Bbb R} \}$ of spatial slices,
which defines, in effect, a time coordinate~$t$ on the spacetime.
(The details of this idea are discussed by York~\cite{York79}, and
familiarity with that material is assumed.) 
The manner in which a spacetime is sliced is a very
important issue when the 3+1~formulation is used as a
basis for performing numerical simulations.
This section recalls some basic ideas on slicing
conditions in numerical relativity and, in particular, describes how
{\it harmonic\/} slicings of spacetimes are constructed.

Traditionally, work in numerical relativity has been based on 
{\it maximal\/} or {\it constant mean curvature\/}~(CMC) slicing
conditions, initially developed for 
this purpose by Eardley, Smarr and York~\cite{York79,SmYo78,EaSm79},
among others. 
A spatial slice has {\it constant mean curvature} if~$K$, the trace of the
extrinsic curvature tensor, takes a constant value on that slice.
If values for~$K$ are specified across a range of slices as a function
of the time coordinate~$t$, then the 3+1~evolution equation for the
extrinsic curvature yields an elliptic equation determining
the value of the lapse function~$N$ on each slice: 
\begin{equation}
  \Delta N - N [ K_{ij} K^{ij} + 4\pi (\rho + S) ]
  = - K'(t)
  \, , \label{Maximal}
\end{equation}
where $\Delta = h^{ij} \nabla_{\! i} \nabla_{\! j}$.
If the function~$K(t)$ is identically zero then the
spacetime is {\it maximally sliced}.
Maximal slicing has been found to work well in numerical simulations
of asymptotically flat spacetimes, while for closed cosmologies (in
which at most one maximal slice can exist) the more general~CMC
slicing condition is appropriate.
Based on their behaviour in simple examples and their success in
numerical simulations, it is believed that the maximal and~CMC
slicing conditions will in general produce foliations
which cover most, if not all, of a spacetime being investigated, and
which at the same time avoid getting too close to any singularities
that may form in that spacetime.

Interest in the use of {\it harmonic\/} slicing has arisen in
recent years because of its connection with work that has been done in
reformulating Einstein's equations as an explicitly {\it hyperbolic\/}
system (see the review~\cite{Reul98} by Reula).
To date, all of the known 3+1~formulations of general
relativity as a strongly hyperbolic evolution system with only
physically relevant characteristic speeds determine the lapse through
some form of the harmonic slicing condition. 

Harmonic slicing contrasts in several basic ways with maximal and~CMC
slicing. 
The lapse function~$N$ in a
harmonically-sliced spacetime is determined either through an
algebraic condition or a simple evolution equation, and not
through an elliptic condition as in equation~(\ref{Maximal}).
This `local' specification of the lapse is a great advantage from the
point of view of implementing a 
numerical code to evolve solutions to Einstein's equations since
elliptic partial differential equations are computationally very
expensive to solve.
The harmonic slicing condition also differs from the maximal and~CMC
slicing conditions in that it specifies a 
relationship between two adjacent slices in a foliation rather than
determining properties of each individual slice: while an isolated
spacetime slice can be characterized as maximal or of constant mean
curvature, there is no such thing as an individual harmonic slice.
One consequence of this is that the choice of lapse function on the
initial slice of a foliation is arbitrary if the harmonic slicing
condition is used.

To simplify the following discussion, all spacetime
foliations considered in this paper are assumed to have shift vectors~$N^i$
which are identically zero.
Since the shift vector controls only the
positioning of spatial coordinates on slices and not how the slices
themselves are arranged, this assumption does not limit the generality
of the results derived here on the appearance of coordinate
singularities. 

The harmonic slicing condition for determining the lapse~$N$ can
be expressed in either algebraic form, 
\begin{equation}
  N = Q(t,x^k) \sqrt{\det h_{ij}}
  \, , \label{HarmFixed}
\end{equation}
or as an evolution equation,
\begin{equation}
  { \partial N \over \partial t } = N (\ln Q)_{,t} - N^2 K
  \, , \label{HarmEvolve}
\end{equation}
where the {\it slicing density}~$Q(t,x^i)$ is an arbitrary
(positive) function of the foliation coordinates which does not depend
on any evolved variables.
If the slicing density is independent of the foliation time coordinate~$t$,
then the harmonic slicing is described as {\it simple}, and in this case a
choice of the value of the lapse~$N$ on an initial slice completely
determines the value of the slicing density~$Q$.
The alternative situation is described as {\it generalized\/} harmonic
slicing, and it is clear from equation~(\ref{HarmFixed}) that any
spacetime foliation can be constructed using generalized harmonic
slicing and a particular form for the slicing density~$Q$.
If the slicing density has only a `separable' dependence on the time
coordinate of the form $Q(t,x^i) = f(t) \bar{Q}(x^i)$, then the
resultant foliation has the same slices as the foliation produced by
using~$\bar{Q}(x^i)$ as the slicing density, but with the slices
labelled by a different time coordinate.
In general it is not obvious how a useful slicing density~$Q$ which
has a non-trivial dependence on the time coordinate~$t$ may be chosen, and
so the simple form of harmonic slicing is most often used in
practice.

The main question addressed in this paper is that of how
suitable the harmonic slicing condition is for numerical work.
Several authors have already considered this question from various
different viewpoints.
Bona and Mass\'o~\cite{BoMa88} have shown that the (simple) harmonic
slicing condition can be used to foliate several standard spacetimes,
and also that `focusing singularities' are avoided by the slicing
condition in much the same way as they are in maximally-sliced
spacetimes, with the slices of the foliation not reaching the
singularity in a finite coordinate time.
(It should be noted that for a harmonically-sliced spacetime a focusing
singularity is essentially a point at which the lapse~$N$ becomes
zero---this is in contrast to the behaviour found at the `gauge
pathologies' described below.)
Cook and Scheel~\cite{CoSc97} have investigated the construction of
well-behaved harmonic foliations for Kerr-Newmann black hole
spacetimes.

The work of Alcubierre and Mass\'o~\cite{Alcu97,AlMa98} is of
particular relevance to the present discussion.
They have shown that `gauge pathologies' (described as `coordinate
shocks' in the earlier paper) can occur in numerical simulations based
on hyperbolic formulations of Einstein's equations which use harmonic
slicing. 
(In fact, a range of gauge conditions are considered by the authors,
with simple harmonic slicing---the only gauge choice of interest
here---corresponding to the special case~$f=1$ in their formulation.
None of the alternative gauge choices they use correspond to the
generalized harmonic slicing condition.)
These gauge pathologies manifest as a loss of continuity at points in
the evolved solution with, in particular, large spikes appearing
in the lapse.
After the time at which the pathologies appear the
numerical solution no longer converges at the expected order.
Alcubierre and Mass\'o explain the appearance of gauge pathologies in
terms of nonlinear behaviour in the hyperbolic evolution equations.
However this fails to adequately answer questions about how common the
gauge pathologies are, what happens to the foliation at the points
where pathologies appear, and what approaches can be used to prevent
the pathologies from occurring.

Simple harmonic slicings for the Schwarzschild and Oppenheimer-Snyder
spacetimes have been investigated by 
Geyer and Herold~\cite{GeHe97,GeHe95}. 
The approach they use is based on an alternative representation of
equations~(\ref{HarmFixed}) and~(\ref{HarmEvolve}) in the
simple harmonic case: if~$T$ is a scalar function on a spacetime, the
level surfaces of 
which represent the spatial slices of a foliation~$\{\Sigma_T\}$, then
the spacetime will be harmonically sliced (in the simple sense) if
\begin{equation}
  \Box T 
  \equiv g^{\mu\nu} \nabla_{\! \mu} \nabla_{\! \nu} T
  = 0
  \, . \label{HarmTime}
\end{equation}
By numerically integrating equation~(\ref{HarmTime}) with respect
to known background metrics, Geyer and Herold construct simple harmonic
slicings which they compare to maximal slicings of the spacetimes.
Some properties of a slicing can be determined straightforwardly from
its time function~$T$, and in particular the lapse~$N$ can be evaluated
through the equation 
\begin{equation}
  g^{\mu\nu} ( \nabla_{\! \mu} T ) ( \nabla_{\! \nu} T )
  = - 1 / N^2
  \, , \label{HarmLapse}
\end{equation}
where it is clear that the vector field normal to the foliation must
remain timelike if the lapse is to have a positive real value.
In fact, Geyer and Herold~\cite{GeHe97} find that for simple harmonic
slicings of the Oppenheimer-Snyder spacetime, foliations which are
initially timelike can at later times become null or spacelike, with
the development of the foliation thus terminating at what it seems
appropriate to call a {\it coordinate singularity}.
The lapse becomes infinite as these singular points are reached, and
this is consistent with the behaviour found at the gauge pathologies
of Alcubierre and Mass\'o. 

In what follows, Geyer and Herold's approach is applied to Minkowski,
Kasner and Gowdy spacetimes, with the intention being to gain a better
understanding of the circumstances under which coordinate
singularities appear, and in particular to determine whether they are a
rare or a common feature of harmonic slicings.

\section{Harmonic Slicings of the Minkowski Spacetime}
\label{sec.Mink}
In the present work, equations~(\ref{HarmTime})
and~(\ref{HarmLapse}) are used to investigate the formation of
coordinate singularities in (simple) harmonic slicings of cosmological
models.
The following approach is employed.
The metric~$g_{\mu\nu}$ of the spacetime being investigated is assumed
to be known with respect to a coordinate system~$(t,x,y,z)$.
An alternative foliation of the spacetime is constructed by taking as
an initial slice one of the constant time hypersurfaces of the
background coordinate system: the foliation time
coordinate~$T(t,x,y,z)$ is given an initial value
\begin{equation}
  T(t_0,x,y,z) = t_0
  \, , \label{HarmInitial}
\end{equation}
for some value~$t_0$ of the coordinate~$t$.
(It is assumed here that the hypersurface $t=t_0$ is spacelike.)
The lapse function of the foliation can be specified arbitrarily on
the initial slice and this determines via equation~(\ref{HarmLapse})
the value of the first derivative of~$T$ away from that slice:
\begin{equation}
  T_{,t} (t_0,x,y,z) = 
    { 1 \over N_0(x,y,z) \sqrt{ - g^{00}(t_0,x,y,z) } }
  \, . \label{HarmInitDeriv}
\end{equation}
Equations~(\ref{HarmInitial}) and~(\ref{HarmInitDeriv})
provide initial data for the wave equation~(\ref{HarmTime}), and
the solution~$T$ describes a new simple harmonic slicing of
the spacetime.

As an example of how equation~(\ref{HarmTime}) can be used to
find coordinate singularities, consider harmonic foliations of the
Minkowski spacetime, written in standard coordinates as
\[
  g_{\mu\nu} dx^{\mu} dx^{\nu} =
    - dt^2 + dx^2 + dy^2 + dz^2
  \, . 
\]
Suppose that a `planar' slicing of the spacetime is constructed such
that the time function~$T$ depends only on the Minkowski
coordinates~$t$ and~$x$.
Then equation~(\ref{HarmTime}) takes the form of the
one-dimensional wave equation
\[
  T_{,tt} = T_{,xx}
  \, ,
\]
which has the general solution
\[
  T(t,x) = f(t+x) + h(t-x)
  \, ,
\]
for arbitrary functions~$f(u)$ and~$h(u)$.
The lapse associated with this time function can be found from
equation~(\ref{HarmLapse}):
\begin{eqnarray*}
  -1/N^2 & = & T_{,x}{}^2 - T_{,t}{}^2 \\
         & = & - 4 f'(t+x) h'(t-x) \, ,
\end{eqnarray*}
and the lapse will be well behaved as long as the function
$f'(t+x)h'(t-x)$ is positive.

For most problems the next step in the analysis would be to use
equations~(\ref{HarmInitial}) and~(\ref{HarmInitDeriv}) to
specify a value for the lapse on an initial slice of the foliation.
However the present case is sufficiently simple that the appearance of
coordinate singularities can be studied without needing to specify an
initial slice.
If the lapse becomes infinite at a point in the foliation then the
function $f'(t+x)h'(t-x)$ must be zero there, and it is clear that the
function must then be zero at all points along a line
$t+x=\hbox{\it constant\/}$ or $t-x=\hbox{\it constant\/}$.
Consequently, any spacelike slice of that foliation must include a
point in it at which the lapse is infinite.
It follows that for simple harmonic slicings of the `planar'
Minkowski spacetime no coordinate singularities will be present, and
the foliation will cover the whole of the spacetime,
provided that the initial slice of the foliation is everywhere
spacelike.
(This result is consistent with the numerical simulations of Minkowski
spacetime reported on by Alcubierre~\cite{Alcu97}.
No gauge pathologies are discovered for `planar' harmonic
slicings---the $f=1$ case---of flat spacetime, although they are found
in the spherically symmetric case.)

In the following sections a similar analysis is performed for more
complicated spacetimes, and coordinate singularities are found to be
much more common than the above result for the Minkowski spacetime
would suggest.

\section{Harmonic Slicings of the Kasner Spacetime}
\label{sec.Kasner}
In this section, equation~(\ref{HarmTime}) is used to construct
simple harmonic foliations of the Kasner spacetime, with the main
point of interest being the question of whether or not coordinate
singularities (gauge pathologies in the terminology of Alcubierre and
Mass\'o~\cite{AlMa98}) form in the slicings.
Later, in section~\ref{sec.Numeric}, these results are compared to numerical
simulations of the spacetime.

The present work uses the axisymmetric Kasner model described by the
metric 
\begin{equation}
  ds^2 = t^{-1/2} ( - dt^2 + dx^2 ) + t \, ( dy^2 + dz^2 )
  \, , \label{AxiKasner}
\end{equation}
for which a cosmological singularity is present at time~$t=0$.
For convenience in the following analysis (and to strengthen the
connection with the Gowdy cosmologies examined in
section~\ref{sec.Gowdy}) a three-torus topology is imposed on the Kasner
spacetime: periodicity is assumed in each of the coordinates~$x$, $y$
and~$z$ over the range~$[0,2\pi)$.

As in the simple example for the Minkowski spacetime given in
the previous section, here `planar' slicings of the Kasner spacetime are
considered for which the time function~$T$ depends only on the
coordinates~$t$ and~$x$.
A straightforward calculation then shows that for the
metric~(\ref{AxiKasner}), equation~(\ref{HarmTime}) for the time
function of a  harmonically-sliced foliation is equivalent to
\begin{equation}
  T_{,tt} - T_{,xx} + t^{-1} T_{,t} = 0
  \, . \label{KasTimeEqn}
\end{equation}
This equation can be solved by decomposing~$T$ spatially as a sum of
Fourier modes (recalling that, since the model is spatially closed,
the function~$T$ must be periodic in the coordinate~$x$) and the
general solution is found to be
\begin{eqnarray}
  T & = & a_0 \ln t + b_0 \label{KasTime} \\
    &   & {} + \sum_{n=1}^{\infty} \bigl[ Z_{+n}\!(t) \cos(nx)
                                        + Z_{-n}\!(t) \sin(nx) \bigr]
          \, , \nonumber
\end{eqnarray}
with
\[
  Z_{\pm n}\!(t) = a_{\pm n} J_0( \vert n \vert t )
                 + b_{\pm n} Y_0( \vert n \vert t )
  \hbox{ \ \ for } n \neq 0
  \, ,
\]
where~$a_{\pm n}$ and~$b_{\pm n}$ are constants, and~$J_\nu$ and~$Y_\nu$
are Bessel functions of the first and second kinds~\cite{AbSt64}.

Equation~(\ref{HarmLapse}) for the lapse function~$N$ of the
foliation can be written in the present case as
\begin{equation}
  \sqrt{t} \, ( T_{,t}{}^2 - T_{,x}{}^2 ) = 1/N^2
  \, , \label{KasLapse}
\end{equation}
and the requirement that the foliation be well behaved can then be
expressed as
\begin{equation}
  T_{,t} > \vert T_{,x} \vert
  \, , \label{KasSing}
\end{equation}
where it is assumed that the orientation of the time function~$T$ of
the foliation is the same as that of the Kasner time~$t$, and thus
that~$T_{,t}$ is positive.
If the time function~$T$ of equation~(\ref{KasTime}) fails to
satisfy the condition~(\ref{KasSing}) at a point, then the
harmonic slicing must have a coordinate singularity there.

The constants~$a_{\pm n}$ and~$b_{\pm n}$ in
equation~(\ref{KasTime}) can be determined by specifying a
value~$N_0(x)$ for the lapse on an initial hypersurface~$t=t_0$ as in
equations~(\ref{HarmInitial}) and~(\ref{HarmInitDeriv}).
In fact, rather than choosing the initial value for the lapse
directly, it is more convenient to choose an
initial value~$I(x)$ for the time derivative of~$T$, with this being
specified as a Fourier sum,
\begin{eqnarray}
  T_{,t}(t_0,x) & = & I(x) \label{KasInitT} \\
                & = & k_0 + \sum_{n=1}^{\infty} 
                      \bigl[ k_{+n} \cos (nx) + k_{-n} \sin (nx) \bigr]
                \, , \nonumber
\end{eqnarray}
where the constants~$k_{\pm n}$ must be such that~$I(x)$ is everywhere
positive.
The value of the lapse on the initial slice can then be determined
through equation~(\ref{HarmInitDeriv}):
\[
  N_0(x) = { 1 \over t_0{}^{1/4} I(x) }
  \, .
\]
The particular time function~$T$ of equation~(\ref{KasTime})
which fits the initial data of equations~(\ref{HarmInitial})
and~(\ref{KasInitT}) is found to be
\widetext
\begin{equation}
  T = t_0 + k_0 t_0 \ln (t/t_0)
    - { \pi t_0 \over 2 } \sum_{n=1}^{\infty}
      \bigl[ J_0(nt) Y_0(nt_0) - Y_0(nt) J_0(nt_0) \bigr]
      \bigl[ k_{+n} \cos(nx) + k_{-n} \sin(nx) \bigr]
  \, .
  \label{KasT}
\end{equation}
\narrowtext

The simplest harmonic foliations of the Kasner
metric~(\ref{AxiKasner}) are the homogeneous ones for which the
time function~$T$ is independent of the coordinate~$x$.
Then, to within a linear rescaling, the function~$T$ equals
$\ln t$, and so the cosmological singularity in the spacetime
occurs in the harmonic foliation only in the limit as~$T$ tends to
negative infinity.
Clearly the foliation covers the entire Kasner spacetime and
contains no coordinate singularities.

The behaviour of homogeneous harmonic slicings is, however, not
representative of the general case.
As an example, figure~\ref{fig.AnalyticT} shows the foliation which
results from setting the lapse on the initial slice to be
\begin{equation}
  N_0(x) = { 1 \over {\textstyle 1 + {1\over2} \sin x} }
  \qquad \hbox{for } t_0 = 1
  \, , \label{KasExamp}
\end{equation}
which corresponds to the choice of non-zero coefficients $k_0=1$
and $k_{-1}=1/2$ in equation~(\ref{KasInitT}).
When equation~(\ref{KasLapse}) is evaluated to determine the
lapse on this foliation, it is found that in some spacetime regions to
the future of the initial slice the quantity~$1/N^2$ is non-positive;
figure~\ref{fig.AnalyticN}, which plots~$1/N^2$, is filled in
where this happens. 
Coordinate singularities must appear in any slices of the foliation
which intersect these regions, and the development of the foliation
cannot proceed beyond a time $T=T_{\text{sing}}\simeq2.45$ with
the lapse~$N$ 
tending to infinity at points as the limiting slice is approached.
In contrast, if the foliation is extended backwards in time from the
initial slice towards the cosmological singularity, then the lapse
appears to be well behaved on all of the slices. 

It can in fact be shown that, in general, harmonic foliations of the
Kasner spacetime always develop coordinate singularities at
sufficiently late (future) times.
If the time derivative of the function in equation~(\ref{KasT}) is
evaluated at a fixed spatial coordinate~$x=x_0$, then
\widetext
\[
  T_{,t}(t,x_0) = { k_0 t_0 \over t }
  + { \pi t_0 \over 2 } \sum_{n=1}^{\infty} n \, c_n(x_0)
      \bigl[ J_1(nt) Y_0(nt_0) - Y_1(nt) J_0(nt_0) \bigr] 
  \, ,
\]
where it is assumed that the value~$x_0$ has been chosen such that at
least one of the values~$c_n$ is non-zero.
For sufficiently large values of the coordinate~$t$, the
approximations~(9.2.5) and~(9.2.6) of reference~\cite{AbSt64}
can be applied to the Bessel functions in this equation.
The result is that
\[
  T_{,t}(t,x_0) \simeq { k_0 t_0 \over t }
    + \sqrt{ \pi \over 2 \, t } t_0 \sum_{n=1}^{\infty} \sqrt{n} \, c_n
        \bigl[ Y_0(nt_0) \cos (nt-{\textstyle{3\over4}}\pi) 
            - J_0(nt_0) \sin (nt-{\textstyle{3\over4}}\pi) \bigr]
  = { k_0 t_0 \over t } + { \Theta(t) \over \sqrt{t} }
  \, ,
\]
\narrowtext
\noindent
where~$\Theta(t)$ is a periodic function which takes both positive and
negative values. 
It follows that, for some sufficiently large value of the
coordinate~$t$, the value of~$T_{,t}$ must be negative at a point.
Since this violates the condition~(\ref{KasSing}), it must
be the case that a coordinate singularity is present in the foliation.

The above argument shows that coordinate singularities must in general
appear in harmonic slicings of an expanding Kasner cosmology but gives
no indication of how much of the spacetime a slicing will cover before
it becomes pathological.
To investigate this, consider the simple case of a time function~$T$
from equation~(\ref{KasT}) which, like the foliation produced by
the initial data~(\ref{KasExamp}), contains only one non-trivial
mode of amplitude~$k$: 
\begin{eqnarray}
  T & = & t_0 + k_0 t_0 \ln (t/t_0) \label{KasSingle} \\
    &   & {} - k { \pi t_0 \over 2 }
          \bigl[ J_0(nt) Y_0(nt_0) - Y_0(nt) J_0(nt_0) \bigr] \sin(nx)
          \, , \nonumber
\end{eqnarray}
where $0 < k < k_0$.
For a fixed value of the Kasner time~$t$ the foliation will be well
behaved (in that condition~(\ref{KasSing}) is satisfied) if and
only if 
\[
  {k_0 t_0 \over t} >
    { k n \pi t_0 \over 2}
    \max_{x}
    \bigl\{ \vert A(t) \cos (nx) \vert - B(t) \sin (nx) \bigr\}
  \, ,
\]
where
\begin{eqnarray*}
  A(t) & = & J_0(nt) Y_0(nt_0) - Y_0(nt) J_0(nt_0) \, , \\
  B(t) & = & J_1(nt) Y_0(nt_0) - Y_1(nt) J_0(nt_0) \, ,
\end{eqnarray*}
and this condition is equivalent to
\begin{equation}
  F(t) \equiv t^2 \bigl[ A(t)^2 + B(t)^2 \bigr]
  < \left( { 2 k_0 \over k n \pi } \right)^2
  \, . \label{KasSingF}
\end{equation}
It is straightforward to show that~$F'(t) \ge 0$.
Furthermore, if the time~$t$ is assumed to be large enough for
approximations~\cite{AbSt64} to be applied to the Bessel functions
in~$A(t)$ and~$B(t)$, then $F(t) \simeq t \, \Pi(t)$ where~$\Pi(t)$ is
a positive, periodic function.
An estimate of the time~$t=t_{\text{sing}}$ at which a coordinate
singularity develops can then be seen to obey
\begin{equation}
  t_{\text{sing}} \propto ( k_0 / k )^2
  \, ,
\end{equation}
and so, for a harmonic slicing which is perturbed from homogeneity by a
single mode of amplitude~$k$, the amount of the spacetime covered by
the slicing becomes infinite as~$k$ tends to zero.

Considering now the behaviour of the slicing in the region of spacetime
between the initial slice and the cosmological singularity, it follows
from equation~(\ref{KasSingF}) that, for the single mode
case~(\ref{KasSingle}), coordinate singularities never appear.
To see this, note that the function~$F(t)$ must satisfy the
inequality at time~$t=t_0$ since the lapse is
required to be well behaved on the initial slice, and since it is
an increasing function of time it must then also satisfy the
inequality at all earlier times.
In the limit as~$t$ tends to zero (while~$x$ is constant), the Bessel
function behaviour is known from equations (9.1.10--13) of
reference~\cite{AbSt64}, and the time function~$T$ must have
the form 
\begin{eqnarray}
  T & \sim & t_0 \bigl[ k_0 + k J_0(nt_0) \cos(nx) \bigr] \ln t
             \nonumber \\
    & & {} + \hbox{(terms that tend to a constant)} 
  \, . \label{KasZero}
\end{eqnarray}
Since the factor multiplying $\ln t$ is always positive, the
value of~$T$ must tend to negative infinity as the cosmological
singularity is approached.
Thus a single mode harmonic slicing will always cover the whole of the
Kasner spacetime to the past of the initial slice.
 
Returning to the general case in which an arbitrary number of modes
are present in the time function~$T$, the above argument can be used
to derive a simple condition on the initial data~(\ref{KasInitT})
which ensures that a slicing is well behaved in the region of
spacetime between the initial slice and the cosmological singularity.
Suppose that equation~(\ref{KasT}) is split up as
\[
  T = t_0 + \sum_{n=1}^{\infty} \left( T^{(+n)} + T^{(-n)} \right)
  \, ,
\]
where
\widetext
\begin{eqnarray*}
  T^{(+n)} & = &
    k_0^{(+n)} t_0 \ln (t/t_0)
    - k_{+n} { \pi t_0 \over 2 } 
      \bigl[ J_0(nt) Y_0(nt_0) - Y_0(nt) J_0(nt_0) \bigr] \cos(nx)
    \, , \\
  T^{(-n)} & = &
    k_0^{(-n)} t_0 \ln (t/t_0)
    - k_{-n} { \pi t_0 \over 2 } 
      \bigl[ J_0(nt) Y_0(nt_0) - Y_0(nt) J_0(nt_0) \bigr] \sin(nx)
    \, , \\
\end{eqnarray*}
\narrowtext\noindent
and the values~$k^{(\pm n)}_0$ are arbitrary subject to the condition
\[
  \sum_{n=1}^{\infty} \left( k_0^{(+n)} + k_0^{(-n)} \right) = k_0
  \, .
\]
The condition~(\ref{KasSing}) that the foliation be free of
coordinate singularities will then certainly be satisfied if
\[
  T^{(+n)}{}_{\!,t} > \vert T^{(+n)}{}_{\!,x} \vert
  \quad \hbox{and} \quad
  T^{(-n)}{}_{\!,t} > \vert T^{(-n)}{}_{\!,x} \vert
  \quad \forall n 
  \, ,
\]
and, as the analysis of the single mode case shows, this will be true
for $0<t<t_0$ if it is true for~$t=t_0$.
The time derivative of each function~$T^{(\pm n)}$ must therefore be
positive on the initial hypersurface $t=t_0$, and for this to be the
case the values~$k_0^{(\pm n)}$ must be chosen such that
\[
  k^{(+n)}_0 > \vert k_{+n} \vert
  \quad \hbox{and} \quad
  k^{(-n)}_0 > \vert k_{-n} \vert
  \quad \forall n 
  \, .
\]
This can always be done if
\begin{equation}
  k_0 > \sum_{n=1}^{\infty} \bigl( \vert k_{+n} \vert 
                                 + \vert k_{-n} \vert \bigr)
  \, , \label{KasSufficient}
\end{equation}
and so any initial value for the lapse which satisfies this condition
must produce a foliation which is free of coordinate singularities to
the past of the initial slice.
In addition to this, if the behaviour of the time function~$T$ is
considered as~$t$ tends to zero, then it is straightforward to show
(in analogy with equation~(\ref{KasZero}) for the single mode case)
that condition~(\ref{KasSufficient}) ensures that~$T$ tends
to negative infinity for all values of~$x$ as the cosmological
singularity is approached, and hence that the foliation covers all of
the spacetime up to the singularity.
The condition~(\ref{KasSufficient}) on the initial lapse is
sufficient but not necessary for the harmonic foliation to be well
behaved to the past of the initial slice.
In fact, experimentation
with different choices of initial data~(\ref{KasInitT}) suggests
that, even when condition~(\ref{KasSufficient}) is not satisfied,
harmonic foliations may never develop coordinate singularities as the
cosmological singularity is approached, although no proof (or
disproof) of this conjecture has yet been constructed.

\section{Harmonic Slicings of Inhomogeneous Cosmologies}
\label{sec.Gowdy}
To summarize the results of the previous section, foliations of the Kasner
spacetime~(\ref{AxiKasner}) based on the simple harmonic slicing
condition~(\ref{HarmTime}) have been investigated under the
assumptions that one slice of the foliation coincides with a
$t=\hbox{\it constant\/}$ hypersurface of the original metric, and that
the foliation is independent of two of the standard spatial
coordinates.
It is found that when the Kasner cosmology is expanding, the slices of
the foliation must eventually develop coordinate singularities (except
when the lapse is initially constant), with
this happening at late times for foliations which are initially close
to homogeneous.
In contrast, when the Kasner cosmology is collapsing, the results
suggest that coordinate singularities never develop in the foliation
and that the slices extend all the way to the cosmological
singularity.
(This is certainly true if the lapse on the initial slice is chosen
such that the coefficients of equation~(\ref{KasInitT}) satisfy
the condition~(\ref{KasSufficient}), and it may in fact be the case
for all reasonable choices of initial lapse.)

The behavioural differences found in harmonic foliations according to
whether the Kasner spacetime is expanding or collapsing can be
explained by considering the relationship between the lapse
function~$N$ and the mean curvature~$K$ of a slice.
In any foliation, the mean curvature provides a measure of the local
convergence of world lines running normal to the spatial slices:
\begin{equation}
  K = - \nabla_{\!\mu} n^{\mu}
    = - \pounds_{\!n} \bigl({ \textstyle \ln \sqrt{\det h_{ij}} }\bigr)
  \, ,
\end{equation}
where~$n^{\mu}$ is the vector normal to the slices, $\pounds_{\!n}$~is
the Lie derivative along that vector, and~$\sqrt{\det h_{ij}}$ is the
spatial volume element.
The value of~$K$ is positive when the foliation world lines are
locally converging, and negative when they are expanding.
The harmonic slicing condition can be written in terms of the
mean curvature of the foliation using
equation~(\ref{HarmEvolve}):
\[
  { \partial (1/N) \over \partial t } = K
  \, ,
\]
where the harmonic slicing is assumed to be simple with zero shift.
It follows that when the mean curvature~$K$ is positive (as it
may be expected to be for slices of a collapsing cosmology) the value
of~$1/N$ will increase, and so the lapse will approach (but usually
not reach) zero.
(The ability of the harmonic slicing condition to avoid `focusing
singularities' at which the lapse vanishes is discussed by
Bona and Mass\'o~\cite{BoMa88}.) 
Conversely, when the mean curvature is negative (as it typically will
be in an expanding spacetime) the value of~$1/N$ will decrease, and if
it reaches zero then a coordinate singularity of the type investigated
in this paper will develop in the foliation.

The obvious extension of the above analysis is to the consideration of
harmonic slicings of spacetimes more general than the Kasner model.
In fact it turns out to be straightforward to extend the results of
section~\ref{sec.Kasner} to a class of inhomogeneous cosmological
models~\cite{Gowd74}: the unpolarized Gowdy spacetimes on~$T^3$ (the
three-torus). 
These spacetimes are vacuum models of
spatially closed cosmologies with planar symmetry, and can be
represented by the metric
\begin{eqnarray}
  ds^2 & = &
  e^{\lambda/2} t^{-1/2}
    ( - dt^2 + d\theta^2 ) \label{GowMetric} \\
  & & {} + e^P t \bigl[ d\sigma^2
           + 2 Q \, d\sigma \, d\delta
           + ( Q^2 + e^{-2P} ) \, d\delta^2 \bigr]
  \, . \nonumber
\end{eqnarray}
The spatial coordinates~$\theta$, $\sigma$ and~$\delta$ range from~$0$
to~$2\pi$.
A cosmological singularity occurs at time~$t=0$ and the model
expands forever through positive values of~$t$.
The functions~$\lambda$, $P$ and~$Q$ depend on the coordinates~$\theta$
and~$\tau$ only and are required by the topology of the model to be
periodic in~$\theta$. 
The vacuum Einstein equations for the metric~(\ref{GowMetric})
reduce to evolution and constraint equations for the variables~$P$,
$Q$ and~$\lambda$ which cannot in general be solved exactly.
Moncrief~\cite{Monc81} has shown that no singularities appear in the
Gowdy~$T^3$ spacetimes for $0<t<+\infty$, and that this coordinate
range describes the maximal Cauchy development of the models.
Thus, no coordinate singularities of the type investigated
in this paper can be present in the standard Gowdy slicing.

If the analysis of the previous section for harmonic slicings of
the axisymmetric Kasner spacetime~(\ref{AxiKasner}) is applied to
the Gowdy metric~(\ref{GowMetric}) (taking the planar symmetry of
the slicing to coincide with the symmetry of the metric) then it is
found that the time function~$T$ must satisfy
\[
  T_{,tt} - T_{,\theta\theta} + t^{-1} T_{,t} = 0
  \, ,
\]
while the lapse~$N$ of the associated foliation is defined through
\[
  e^{-\lambda(t,\theta)/2} \sqrt{t} 
  \bigl( T_{,t}{}^2 - T_{,\theta}{}^2 \bigr) 
  = 1/N^2
  \, .
\]
Comparing these expressions to equations~(\ref{KasTimeEqn})
and~(\ref{KasLapse}), it is clear that the Gowdy~$T^3$ spacetimes admit
the same harmonic foliations as the axisymmetric Kasner spacetime, and
furthermore that the condition that foliations must satisfy to be free
of coordinate singularities is the same.
It thus turns out that all of the results summarized above on the
behaviour of harmonic slicings of a Kasner cosmology carry over
directly to a fairly general class of planar cosmological models.
(In addition, it may be noted that the class of Gowdy spacetimes
on~$T^3$ includes as special cases the full set of Kasner spacetimes
with the same topology. 
Thus the restriction of the discussion in section~\ref{sec.Kasner} to
only the axisymmetric Kasner model~(\ref{AxiKasner}) can be relaxed;
the same results hold for all of the Kasner spacetimes.)

\section{Numerical Simulations of the Kasner Spacetime}
\label{sec.Numeric}
The results of section~\ref{sec.Kasner} show that coordinate
singularities are a generic feature of harmonic foliations of the
(expanding, axisymmetric) Kasner cosmology.
It follows then that numerical simulations of that spacetime which use
the harmonic slicing condition will be forced to terminate after a
finite number of time steps, regardless of their accuracy.
The practical details of this are examined in the present section.

\begin{figure}
  \begin{center}
    \epsfig{width=8.5cm, file=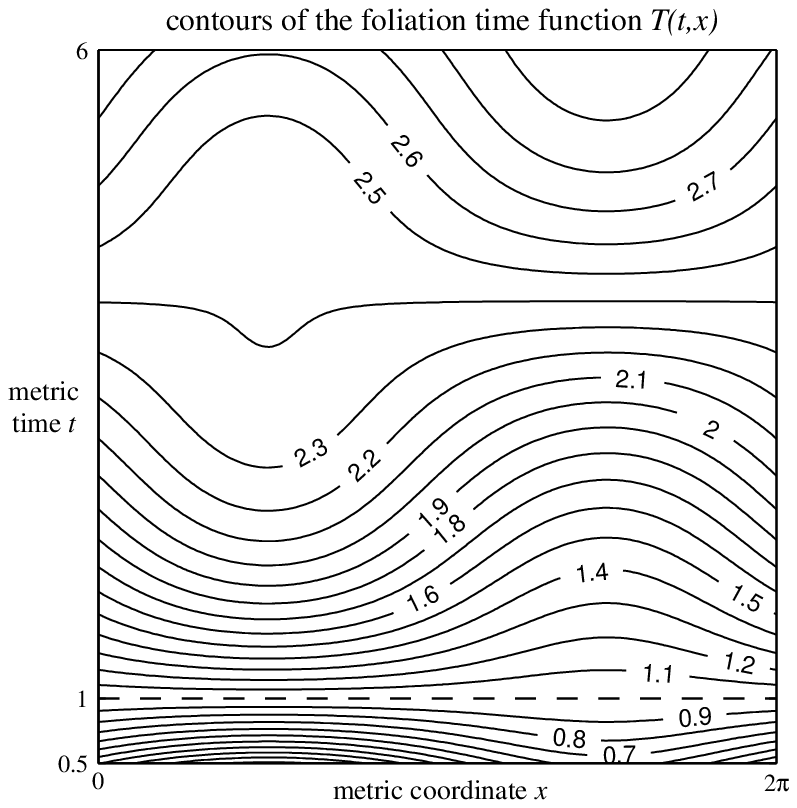}
  \end{center}
  \bigskip
  \caption{A simple harmonic foliation as seen in the background
  Kasner metric. 
  The initial slice of the foliation coincides with the $t=1$
  hypersurface of the metric~(\ref{AxiKasner}) and is given
  an inhomogeneous initial lapse~(\ref{KasExamp}).
  The time function~$T$ is evaluated from equation~(\ref{KasT}) and
  presented as a  contour plot with respect to the
  coordinates~$(t,x)$ of the original metric. 
  The contours represent slices of the foliation with
  the dashed line marking the position of the initial slice.}
  \label{fig.AnalyticT}
\end{figure}

\begin{figure}
  \begin{center}
    \epsfig{width=7.801cm, file=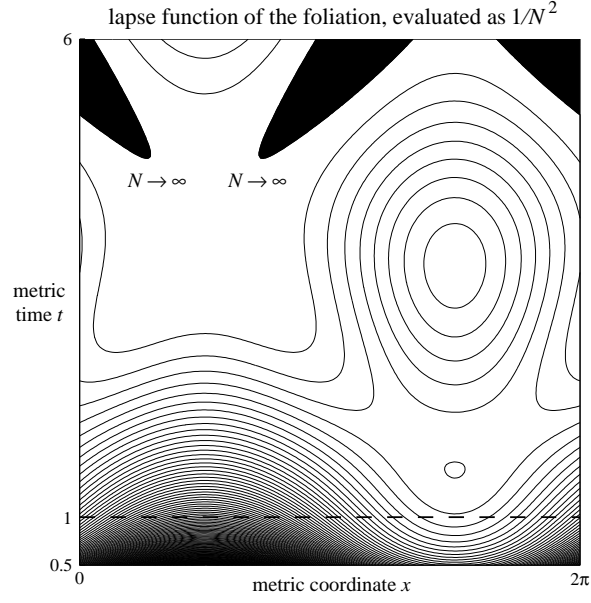}
  \end{center}
  \bigskip
  \caption{Lapse function for the harmonic foliation of
  figure~\ref{fig.AnalyticT}. 
  The lapse~$N$ is evaluated from the foliation time~T through
  equation~(\ref{KasLapse}) and presented with respect to the
  coordinates~$(t,x)$ of the background Kasner metric~(\ref{AxiKasner}). 
  Evenly spaced contours of the function~$1/N^2$ are shown.
  The filled-in regions at the top of the plot show where~$1/N^2$ ceases
  to be a positive function, and thus where coordinate singularities
  must appear in the slices of the foliation.
  As in figure~\ref{fig.AnalyticT}, the dashed line marks the position
  of the initial slice.} 
  \label{fig.AnalyticN}
\end{figure}

Numerical simulations are performed based on initial data for the
Kasner spacetime given by the metric~(\ref{AxiKasner}) on an initial
slice at time $t=t_0=1$ using the
harmonic slicing condition~(\ref{HarmFixed}) and the initial
value~(\ref{KasExamp}) for the lapse.
The foliation of the Kasner spacetime which results is the one
pictured in figure~\ref{fig.AnalyticT}, and the numerical simulation
thus allows the prediction made in section~\ref{sec.Kasner}---that
coordinate singularities prevent the foliation from developing beyond a
time $T=T_{\text{sing}}\simeq2.45$---to be verified.
The slicing density of the foliation is given the inhomogeneous form
\begin{equation}
  Q(T,X,Y,Z) = { t_0{}^{-3/4} \over 1 + {1\over2} \sin X }
  \, , \label{NumDensity}
\end{equation}
where $(T,X,Y,Z)$ is the coordinate system of the numerical
simulation, and is distinct from the coordinate system $(t,x,y,z)$ of
the original Kasner metric except on the initial slice which is
labelled~$T=t_0$ and on which $(X,Y,Z)=(x,y,z)$.
The shift vector~$N^i$ is taken to be identically zero,
and the lack of dependence of the slicing density~$Q$ on the foliation
time~$T$ means that the harmonic slicing is of the simple type.
All of the quantities involved in the numerical simulation are
independent of the spatial coordinates~$Y$ and~$Z$ so that the evolved
spacetime in effect has planar symmetry.
The numerical simulation is one dimensional and uses
a high-resolution numerical scheme within an adaptive mesh
refinement code to evolve a first-order hyperbolic formulation of the
vacuum Einstein equations; a complete description of the
code is given in reference~\cite{Hern99}.

Figure~\ref{fig.Slices} displays the~$(T,X)$ coordinate system of the
numerical simulation as it appears in the~$(t,x)$ coordinates of the
homogeneous background metric~(\ref{AxiKasner}).
This comparison between coordinate systems is made using an algorithm,
described in reference~\cite{Hern99}, for tracking the
positions of the observers at rest in the slices of the
foliation, based on the values~$N(T,X)$ taken by the lapse
during the course of the simulation.
The observer world lines are plotted up until a simulation
time~$T=2.3$.
Although the inaccuracies in the numerical solution
which are present at that time are not sufficient to stop the
simulation, they do make it increasingly difficult to accurately track
the positions of the observers.
The $T=\hbox{\it constant\/}$ surfaces reconstructed in
figure~\ref{fig.Slices} can be seen to be in good agreement with the
exact results shown in figure~\ref{fig.AnalyticT}. 

The analysis of section~\ref{sec.Kasner} (as presented in
figures~\ref{fig.AnalyticT} and~\ref{fig.AnalyticN}) shows that at a
time $T=T_{\text{sing}}\simeq2.45$ the lapse~$N$ must become infinite
at points of the evolved foliation. 
Figure~\ref{fig.LapseCoord} shows the actual
value taken by the lapse at the earlier time~$T=2.3$ in the
simulation, and it can be seen that two sharp spikes are already
present in the solution.
As the evolution progresses these spikes grow in size, and if the
evolved solution were exact they would reach infinite heights within a
short time.
As they grow, the two spikes also narrow and become closer together.
The reason for this can be seen from the paths of the coordinate
observers plotted in figure~\ref{fig.Slices}: in the
region~$0<x<\pi$ where the coordinate singularities eventually form, 
the world lines normal to the foliation are diverging (with respect to
the background coordinates), and so the region is resolved by a
diminishing number of grid points.
It should be noted that this divergence of observers is not itself the
cause of the coordinate singularities; it could in principle be
counteracted by an appropriate choice of shift vector for the
foliation, but this would not prevent the lapse from becoming
infinite.
The narrowing and effective coalescence of the spikes is problematic
for the simulation since features which are smaller than the grid
spacing~$\Delta X$ cannot be accurately resolved, and, regardless of
the number of grid points used, eventually the numerical solution
must fail to accurately model the behaviour of the exact solution.
In fact, with the spikes being inadequately resolved, the evolved
value of the lapse fails to become infinite (or even the computer
representation of this) as the coordinate singularities are reached, and
it is possible for the simulation to continue beyond the time at which
the slices of the foliation cease to be spacelike in the exact solution.
However, the numerical solution has no physical meaning past the
points at which coordinate singularities form, and in particular, as
noted by Alcubierre and Mass\'o~\cite{AlMa98}, it will no longer 
converge as the grid is refined. 

The spikes that develop in the lapse as the coordinate
singularities are approached are at face value very similar to some
features that are seen in numerical solutions for collapsing
one- and two-dimensional inhomogeneous cosmologies, as studied by 
Hern and Stewart~\cite{Hern99,HeSt98}.
In both cases spiky features appear in the evolved variables which, as
the simulation progresses, increase in height and decrease in width
until they can no longer be resolved by the numerical grid used in the
simulation.
The question then arises as to whether the features seen in the
inhomogeneous cosmologies have any physical relevance or whether they
are, like the features in the Kasner simulation described above,
entirely a coordinate effect.

In reference~\cite{Hern99} it is concluded that the spiky features
found in the inhomogeneous cosmological models are not simply
artifacts of the coordinate systems (or the metric variables) on which
the simulations are based.
Part of the evidence for this comes from an examination of the
profiles of the evolved variables with respect to proper distances
rather than coordinate distances: the narrowing of the spiky features
becomes even more pronounced when proper distances are used.
It is interesting to note that the opposite result is observed when 
a similar analysis is applied to the spikes seen in the lapse in
figure~\ref{fig.LapseCoord}.
If the metric component~$h_{11}$, which measures the proper distance in
the $X$-direction of the simulation, is examined for the data plotted
in figure~\ref{fig.LapseCoord}, it is found that sharp spikes
(towards positive infinity) are coincident there with the spikes in
the lapse, and hence that the apparently narrow spikes are in fact
spread over large physical distances. 
As figure~\ref{fig.LapseProper} demonstrates, 
if instead of being plotted against the coordinate~$X$, the lapse is
shown as a function of the proper distance along the $X$-axis, then
the spikes no longer appear as notable features of the solution.
This approach could prove to be a useful way of distinguishing genuine
physical features in an evolved solution from effects caused by
coordinate singularities in harmonic slicings.

\section{Discussion}
\label{sec.Discuss}
It has been demonstrated in sections~\ref{sec.Kasner}
and~\ref{sec.Gowdy} of this paper that `planar' foliations of Kasner
and Gowdy~$T^3$ cosmologies generated using the simple harmonic
slicing condition will, except in special cases, always terminate at
coordinate singularities without covering the whole of the spacetimes.
When these analytic results are considered together with the
descriptions of coordinate singularities in specific examples of
`spherically symmetric' harmonic foliations given by
Alcubierre~\cite{Alcu97}, and Geyer and Herold~\cite{GeHe97}, a
pessimistic picture  of the nature of the harmonic slicing condition
emerges: 
it appears likely that, except in special cases, foliations generated by
harmonic slicing will eventually terminate at coordinate singularities. 
(Seemingly this pathological behaviour can occur whenever the
foliation locally undergoes a protracted period of expansion).

The coordinate singularities examined here arise as part of exact
solutions for harmonically-sliced spacetimes, and they cannot be
avoided simply by employing alternative numerical methods in
simulations.
This is clearly a drawback to the use of harmonic slicing in
numerical work, and by extension to the use of hyperbolic
formulations of Einstein's equations which typically rely on this type
of slicing. 
It should be appreciated however that no known slicing condition is
guaranteed to completely cover an arbitrary spacetime, and potential
problems with coordinate singularities do not necessarily outweigh the
advantages of using hyperbolic formulations in numerical relativity.
In section~\ref{sec.Numeric} the behaviour of a numerically evolved
solution is examined as it approaches a coordinate singularity, and it
is found that by considering measures of proper distance it may be
possible to distinguish the effects of coordinate singularities from
genuine physical features of a spacetime.
In general though the most reliable approach to identifying the
presence of a coordinate singularity in a numerical solution is the
one recognized by Alcubierre and Mass\'o~\cite{AlMa98}: convergence of
the solution is lost when a coordinate singularity develops.

\begin{figure}
  \begin{center}
    \epsfig{width=8.5cm, file=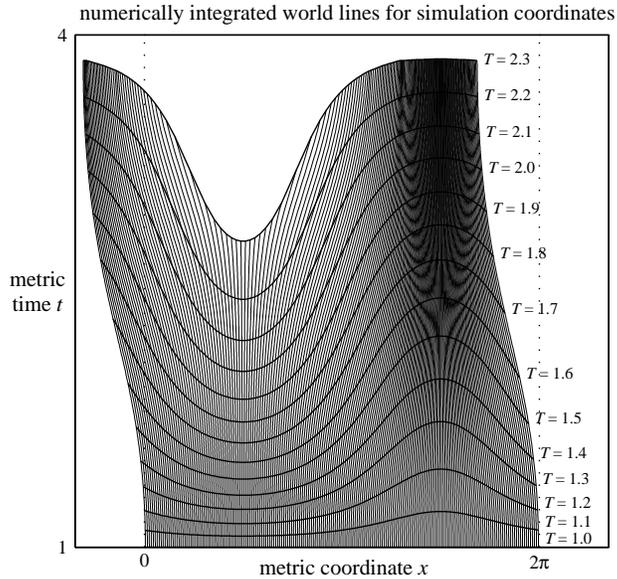}
  \end{center}
  \bigskip
  \caption{Coordinate world lines for a numerical simulation as seen
  in the background Kasner metric.
  A homogeneous initial data set is numerically evolved using harmonic
  slicing based on the inhomogeneous gauge function~(\ref{NumDensity}).
  The world lines of the simulation's grid points are tracked relative
  to the coordinate system~$(t,x)$ of the background
  metric~(\ref{AxiKasner}) using the numerical values produced for the
  lapse function. 
  Two hundred grid points are evenly spaced on the $t=1$
  hypersurface of the background metric at the initial simulation
  time $T=1$, and the paths they subsequently follow are plotted up
  until the end of the simulation at time~$T=2.3$.
  Periodicity of the spatial domain means that observers who `go off'
  at $x=0$ will `come on' at $x=2\pi$.
  This plot is to be compared with figure~\ref{fig.AnalyticT}.}
  \label{fig.Slices}
\end{figure}

\begin{figure}
  \begin{center}
    \epsfig{width=8.5cm, file=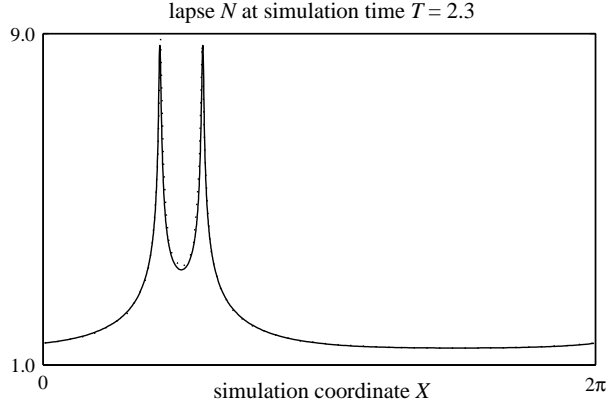}
  \end{center}
  \bigskip
  \caption{Lapse function for the inhomogeneously-sliced Kasner
  spacetime. 
  The solid line shows the value of the lapse~$N$ at the final time of
  the numerical simulation presented in figure~\ref{fig.Slices}. 
  Coordinate singularities are responsible for the sharp spikes in the
  data. 
  (The dotted line shows an alternative value for the lapse evaluated
  indirectly by combining the estimated world line positions of
  figure~\ref{fig.Slices} with the analytic lapse values plotted in
  figure~\ref{fig.AnalyticN}; the results of both approaches can be
  seen to be in good agreement.)}  
  \label{fig.LapseCoord}
\end{figure}

\begin{figure}
  \begin{center}
    \epsfig{width=8.5cm, file=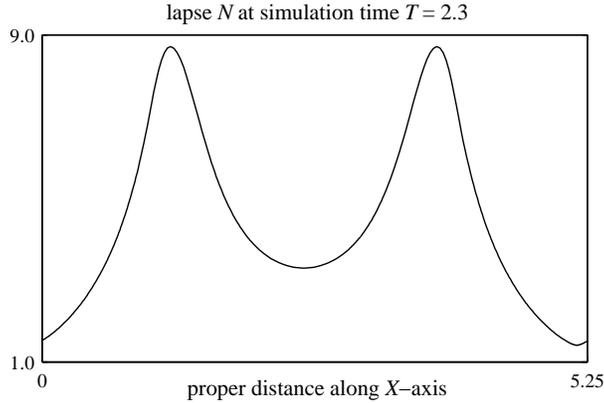}
  \end{center}
  \bigskip
  \caption{The lapse from figure~\ref{fig.LapseCoord} is re-plotted
  against proper distance in the $X$-direction.
  The peaks which in the original plot appear very sharp are found to be
  quite smooth when viewed with respect to a physical measure of
  distance.
  (The proper distance is measured starting from the point $X=0$, and an
  orbit of the evolved spacetime at time $T=2.3$ is found to have a
  proper length of approximately~$5.25$.)}
  \label{fig.LapseProper}
\end{figure}

Of course, the formation of coordinate singularities has only been
discussed here for the simple harmonic slicing condition, and it is
possible that the freedom in choosing the time dependence of the
slicing density in generalized harmonic slicing could be put to use in
controlling the development of the spacetime foliation such that
coordinate singularities are avoided. 
Although when used in the context of a hyperbolic formulation the slicing
density is formally required to be independent of the evolved
variables, in practice there seems to be no problem in allowing
occasional `corrections' to be made to its value in response to the
behaviour of the foliation; in effect this amounts to intermittently
halting the simulation and choosing a new value for the lapse on the
current slice.
(An important point here is that changes to the slicing density should
be made only at fixed time intervals, rather than after a fixed number
of time steps, since otherwise the exact solution being sought will
depend on the resolution of the simulation.)
This `piecewise harmonic' form of slicing has been used in some
preliminary tests employing simple heuristics for making alterations to
the value of the slicing density, with the resulting foliations being
examined using the world line integration algorithm which was used to
produce figure~\ref{fig.Slices}.
(The algorithm was in fact developed for this purpose.)
As yet however no definite conclusions regarding the
effectiveness of this approach have been reached.

\acknowledgements

Thanks must go to John Stewart for overseeing this research.
The author was supported by a studentship from the
Engineering and Physical Sciences Research Council.

\newpage

\newpage

\newpage

\newpage

\end{document}